\documentstyle[12pt,version2,aps]{revtex}

\begin{document}

\begin{title}
 Quantum Group and Magnetic Translations.\\
 Bethe-Ansatz for Asbel- Hofstadter Problem\\
\end{title}
   \author{P.B Wiegmann* and A.V.Zabrodin**}

\begin{instit}

* James Frank Institute and Enrico Fermi Institute of the University
of
Chicago,\\
5640 S.Ellis Ave.,Chicago Il
60637,e-mail:WIEGMANN@CONTROL.UCHICAGO.EDU
\\
\\Landau Institute for Theoretical Physics\\
** Institute of Theoretical Physics, Uppsala University,\\
Box 803 S-75108, Uppsala, Sweden\\
and \\the Institute of Chemical Physics, Kosygina St. 4, SU-117334,
Moscow,
Russia \end{instit}
    \begin{abstract}

We present a new approach to the problem of Bloch electrons in
magnetic ( sometimes called Azbel-Hofstadter problem)
field, by making explicit a natural relation between the group of magnetic
translations
and the quantum group $U_{q}(sl_2)$.  The approach
allows us to express the "mid" band spectrum of the model and the
Bloch wave function as solutions of the Bethe-Ansatz equations
typical for completely integrable quantum systems. The zero mode wave
functions are found explicitly in terms of
$q$-deformed classical orthogonal polynomials.In this paper we present solution
for the isotropic problem. We also present a class of solvable
quasiperiodic equations related to  $U_{q}(sl_2)$.
\\
\end{abstract}

%\pacs{PACS numbers: 71,75, 02}

%\begin{narrowtext}

\section{Introduction}

The peculiar problem of Bloch electrons in magnetic field
\cite{Z},\cite{A},\cite{W},\cite{H},\cite{TKNN} often emerges in various
branches physics. Every time presenting a new face to
describe another physical application.
The Hamiltonian of a particle on a two dimensional square lattice in
magnetic field is
\begin{equation}
H =  \sum_{<n,m>} t_{n,m}
e^{iA_{\vec{n},\vec{m}}}c_{\vec{n}}^{\dag}c_{\vec{m}},
\label{ham}
\end{equation}
\begin{equation}
\prod_{plaquette}e^{iA_{\vec{n},\vec{m}}}=e^{i\Phi}
\end{equation}
where $\Phi=2\pi{P\over Q}$ is a flux per plaquette, $P$ and $Q$
are
mutual prime integers and $t_{n,m}$ is a hoping amplitude between the nearest
neighbors.
In the most conventional Landau gauge
$A_{x}=A_{\vec{n},\vec{n}+\vec{1}_x}=0,
A_y=\Phi n_{x}$ the  Bloch wave function is
\begin{equation}
\psi(\vec n)=e^ {i\vec k \vec n}\psi_{n_x}(\vec
k),\,\,\,\,\psi_{n}=\psi_{n+Q}
\label{psi}
\end{equation}
where $n_{x}\equiv n=1,...,Q$ is a coordinate in the magnetic cell.
With
these substitution the Schrodinger equation  turns into a  famous
one-dimensional  quasiperiodic difference equation  ("Harper's"
equation):
\begin{eqnarray}
\label{harper}
 t_x (e^{ik_x}\psi_{n+1}+e^{-ik_x}\psi_{n-1})
+2t_y \cos(k_{y}+n\Phi)\psi_n=E\psi_n
\end{eqnarray}
The spectrum of this equation has $ Q$ bands and feels the difference
between rational and irrational numbers \cite{A}. The beuaty and complexity of
this problem is the hierarchy of the spectrum. If the flux is irrational,
the spectrum
is singular continuum - uncountable but measure
zero set of points (Cantor set) \cite{H}. The{\it multifractal} behaviour  (
see
\cite{K} for a review) of the spectrum is a long standing theoretical chalenge.

The Harper's equation (i.e the Shrodinger equation
for the Bloch particle in the Landau gauge) formally describes one dimensional
electrons in quasiperiodic potential and also one dimensional quasicristal.
 Depending of the stength of the potential $t_x/t_y$ (it corresponds to
anisotropy in hopping amplitude of a particle in magnetic field) the Harper
equation describes localization- delocalization transition in one dimensional
incommensurate potential.

The list of others applications  may be continued.

Recently it has been
conjectured that the symmetry of magnetic group may appear dynamically in
strongly correlated electronic systems\cite{Anderson},\cite{wiegmann}.

In this paper we made explicit a long time anticipated connection  of the
noncommutatative geometry in magnetic field and the structure of the
 {\it Quantum Group} $U_{q}(sl_2)$ and therefore with Quantum
Integrable Systems \cite{wieg}. We present the {\it Bethe- Ansatz}
 algebraic equations for
the  spectrum of isotropic problem. Although we do not solve the Bethe-Ansatz
equations here, we hope that they provide a basis for analytical studying the
fractal properties of the spectrum.
Although, the Bethe Ansatz solution is available for an anisotropic case, in
this paper we present only isotropic solution $t_x=t_y=1$. Anisotropic case wil
be considered elsewhere.

The spectrum of our problem is complex but not chaotic. Instead it is governed
by the Quantum integrability. It may happened that the Bethe Ansatz technique
will be a powerfool tool to study general properties of multifractality. In
fact multifractality has been already observed in quantum integrable models
\cite{Tak}, \cite{Jap}. However, this aspect of integrability has never been
seriously developed.

Let us stress that we apply the Bethe Ansatz
technique to Quantum Mechanics - not to a Quantum Field Theory. Therefore we
found convinient to use the theory of representation of the Quantum Group,
instead of traditional Bethe Ansatz techique.

The Harper's equation is a member ( perhaps the most interesting but at least
the most famous) of class of solvable quasiperiodic equations related to
$U_q(sl_2)$. It will be briefly described below ( for a more extended
discussion
of algebraization of quasiperiodic equations see Ref.\cite{Zab}).

To ease the reference we state the main result now:

It is known that due to the gauge invariance, the energy depends on a
single parameter $\Lambda= \cos (Qk_x)+\cos (Qk_y)$ \cite{T}. We find that the
spectrum at $\Lambda= 0 $ ( "mid" band spectrum ) is given by the sum
of
roots $z_l$\begin{equation}
E=iq^Q(q-q^{-1})\sum_{l=1}^{Q-1}z_l,\label{energy1}
\end{equation}
of the {\it Bethe-Ansatz } equations for the Quantum Group
$U_q(sl_2)$ with

\begin{equation}
{{z_l^2+q}\over {qz_{l}^2 +1}}=-
q^Q\prod_{m=1,m\ne l}^{Q-1} {{q z_l-z_m}\over {z_{l}-q z_m}},\,\,\,
\,l=1,...,Q-1.
\label{be}
\end{equation}
\begin{equation}
 q=e^{{i\over 2}\Phi},
\label{q}
\end{equation}
Another version of the Bethe-Ansatz equations is presented in Sect.7.
An ambitious problem of extending the Bethe Ansatz solution forarbitrary
momenta $k_x,k_y$ is beyond of the scope of this paper.
 The paper is organized as follows. In Sect.2 we present the Schrodinger
equation in two
other gauges in which the quantum group structure of the model is
more transparent. In Sect.3 we review some necessary facts about
the quantum group $U_q(sl_2)$ and its representations.
The relation between the quantum group and magnetic translations
(\ref{tr}) is revealed in Sect.4
The Hamiltonian (\ref{ham}) can be rewritten
entirely via the quantum group generators . We give two equivalent ( dual)
representations of
the Hamiltonian as a  linear or quadratic form in the
quantum group generators. In Supplement A we consider a general
quadratic forms in
$U_q(sl_2)$
generators and introduce an "integrable " class of the second order
 "quasiperiodic"
equations  related to the quantum group. In Supplement B we show that a
general quadratic form in quantum group generators is atrace of monodromy
matrix of some integrable models with nonperiodic boundary conditions.
In Sect.5
we apply the functional Bethe Ansatz
 to obtain the "mid" band spectrum of the model. A
promising connection between the zero mode wave functions and $q$-deformations
of certain classical orthogonal polynomials is discussed in Sect.6.

\section { Magnetic Translations  and  gauges invariance}
The wave functions of a particle in a magnetic field form a
representation of
the
{\it group of magnetic translations} \cite{Z}: let generators of the
translations be
\begin{equation}
T_{ \vec \mu} (\vec i)=e^{iA_{\vec i,\vec  i+\vec \mu}}\mid \vec
i><\vec  i+\vec \mu\mid
\end{equation}
They form the algebra
\begin{eqnarray}
\label{tr}
&&T_{\vec{\mu}}=T_{-\vec{\mu}}^{-1},\,\,\,
T_{\vec{n}}
T_{\vec{m}}=q^{-\vec{n}\times \vec{m}}T_{\vec{n}+\vec{m}},
\nonumber\\
&&T_yT_x=q^2gT_y, \,T_yT_{-x}=q^{-2}T_{-x}T_y
\end{eqnarray}
with $q=\exp{i\pi{P\over Q}}$.
The Hamiltonian (\ref{ham})
therefore
is
\begin{equation}
H=T_x+T_{-x}+T_y+T_{-y}
\label{ham2}
\end{equation}

Integrability of the problem can be seen from the first glance. It has been
observed (see e.g.\cite{T}) that the spectrum depends only on one parameter
$\Lambda$. That means that there is a parametric family of
 Hamiltonians
\begin{equation}
H(u)=uT_x+u^{-1}T_{-x}+vT_y+v^{-1}T_{-y} \label{com}
\end{equation}
with the same spectra at
$$u^Qe^{ik_x}+u^{-Q}e^{-ik_x}+v^Qe^{ik_y}+v^{-Q}e^{-ik_y}=const$$.
 The mid band spectrum $\Lambda$=0 is a very spetial point where the
Bethe Ansatz is especially simple. We shall consider only this point below. The
useful way to explore this idea is representation teory of quantum groups.

We found  two specific gauges for which the
quantum group structure is more transparent. In these gauges (we call them
"regular") the wave function has the form
\begin{equation}
\Psi_n=\prod_{m=1}^{Q-1} (q_{0}^{n}-z_m)
\label{polynom1}
\end{equation}
where  $q_{0}=q$ or $q_{0}=q^2$ and
  $z_m$'s are the  roots of the Eq.(8) and are independent on $n$.
The Landau gauge of the Introduction is not "regular"
in this sense and is not convenient for
revealing the quantum group structure of the model.

{\bf 1.Modified Landau gauge}

Consider the gauge
\begin{equation}
\label{MLG}
A_{x}=-A_{y}=-\Phi n_{x}
\end{equation}
which we refer as a modified Landau gauge. Now he Bloch wave function
is \begin{equation}
\phi(\vec n)=e^ {i\vec k' \vec n}\phi_{n_x}(\vec k'
),\,\,\,\,\phi_{n}=\phi_{n+Q}
\label{phi}
\end{equation}
where $\vec k'$ is the wave vector (different from that in
(\ref{psi})). The Schrodinger equation turns in to
\begin{eqnarray}
\label{modharper}
 e^{i k'_{x}-i\Phi n}\phi_{n+1}+e^{-i k'_{x}+i\Phi (n-1)}\phi_{n-1}
+2\cos( k'_{y}+n\Phi)\phi_n = E\phi_n
\end{eqnarray}
(we drop the argument in $\phi_{n}$).

The "mid" band spectrum ($\Lambda =0$) corresponds to the values
\begin{equation}
\label{mid1}
\vec k=({{1}\over {2}}\Phi (Q-1), \pi )
\end{equation}
in the Landau gauge and is given by
\begin{equation}
\label{mid2}
 \vec k'=(0,\pi )
\end{equation}
in the Modified Landau gauge.

At the "mid" band all the values of $\vec k'$ on the line
$\exp{i(k'_x-k'_y)}=-1$ are physically equivalent i.e. connected by a gauge
transformation. We have chosen $\exp{i(k'_x+k'_y)}=-1$ in (\ref{mid2}).

{\bf 2.Chiral gauge}

Consider the chiral gauge defined by:
\begin{equation}
\label{LC}
A_{x}=-{{\Phi}\over {2}}(n_{x}+n_{y}),\,\,\,\,\,\,
A_{y}={{\Phi}\over {2}}(n_{x}+n_{y}+1),
\end{equation}
in wich the Bloch wave function takes the form
\begin{equation}
\chi(\vec n)=e^ {i\vec p \vec n}\chi_{n}(\vec p)
\label{chi}
\end{equation}
where $\vec p=(p_{x},p_{y}), n=n_{x}+ n_{y},
p_{\pm}=(p_{x}\pm p_{y})/2$ are light cone coordinates and momenta.

This wave function is defined in {\it two} magnetic cells because
$n$ runs now from 1 to $2Q$. Accordingly, the light cone momenta are
confined to the half of the Brillouin zone $[0,{\pi}/Q]$. The
equivalent
form of the Harper's equation (\ref{harper}) is
\begin{eqnarray}
\label{LCharper}
&& 2e^{{i\over 4}\Phi + ip_{+}} \cos ({1\over 2}\Phi n+ {1\over
4}\Phi
-p_{-})\chi_{n+1} +
\nonumber\\
&&2e^{-{i\over 4}\Phi - ip_{+}} \cos ({1\over 2}\Phi n- {1\over
4}\Phi
-p_{-})\chi_{n-1}
=E\chi_{n}
\end{eqnarray}
The  wave function $\chi_{n}$ is $2Q$-periodic. This doubling of
period
in comparison with (\ref{modharper}) is, of course, artificial.
Although
the coefficients in (\ref{LCharper}) have period $2Q$ there exists a
simple transformation of $\chi_{n}$ which makes them $Q$-periodic:
\begin{equation}
\label{trans2}
\chi_{n}=\exp({i\Phi \over 4}n(n-2))\xi_{n}
\end{equation}
This new wave function $\xi_{n}$ is $Q$-periodic and satisfies the
equation
\begin{eqnarray}
\label{modLCharper}
&& (e^{ -{i\over 4}\Phi + ip_{x} }+e^{i\Phi n+ {i\over 4}\Phi
+ip_{y} })\xi_{n+1} +
\nonumber\\
&&(e^{ {i\over 4}\Phi - ip_{x} }+ e^{-i\Phi n + {3i\over 4}\Phi
-ip_{y}})\xi_{n-1}
=E\xi_{n}
\end{eqnarray}
The "mid" band spectrum corresponds to $\vec p=({1\over 2}\pi ,{1\over
2}\pi+{1\over  4}\Phi)$ in the chiral gauge (compare with
(\ref{mid2}) ). Again all the values $\vec p =({1\over 2}\pi , \delta)$ for any
real $\delta $ are physically equivalent. We choose $\vec p$ at the "mid" band
to be
\begin{equation}
\label{mid3}
\vec p =({{\pi}\over {2}},
{{\pi}\over {2}})
\end{equation}
The relation between the three gauges are given in the Appendix A.

\section {Quantum Group}

The algebra $U_q(sl_2)$ ( a q-deformation of the universal enveloping
of the $sl_2$)
is generated by the elements $A,B,C,D,$ with the commutation
relations
\cite{KR},\cite{Sk1},\cite{Dr},\cite{J},\cite{FRT}
\begin{eqnarray}
\label{ABCD}
&&AB=qBA,\,BD=qDB,
\nonumber\\
&&DC=qCD,\,CA=qAC,
\nonumber\\
&&AD=1,\,[B,C]={{A^2-D^2}\over{q-q^{-1}}}
\end{eqnarray}
 We shall take  the deformation
parameter $q$ as a root of $\pm 1$ of degree $Q$:
$q=exp(i\pi{P\over Q})$ where $P$ and $Q$ are mutually prime
integers.
The central element
of this algebra (for arbitrary $q$) is a $q-$analog of the Casimir
operator
\begin{equation}
c_0=\left({{q^{-{1\over 2}}A-q^{1\over 2}D}\over {q-q^{-1}}}\right)^2
+BC
\label{casimir}
\end{equation}
When $q$ is a root of unity some additional central elements appear.

In the classical limit $q\rightarrow {1+{i\over 2}\Phi}$, the quantum
group
turns to the $sl_2$ algebra:  ${(A-D)/(q-q^{-1})}\rightarrow S_3,\,
B\rightarrow S_{+},\,C\rightarrow S_{-},\,c\rightarrow {\vec S}^2
+1/4$.

The commutation relations (\ref{ABCD}) are simply another way to
write the
interwining relation for the $L$ - operator:
\begin{equation}
R({u/v})(L(u) \otimes 1)(1 \otimes L(v))=
(1 \otimes L(v))(L(u) \otimes 1)R({u/v})
\label{RLL=LLR}
\end{equation}
with the trigonometric $R-$matrix
\begin{equation}
R(u)={1\over 2}(q+1)(u-q^{-1}u^{-1})+
{1\over 2}(q-1)(u+q^{-1}u^{-1})\sigma_{3}\otimes \sigma_{3}+
(q-q^{-1})(\sigma_{+}\otimes \sigma_{-}+\sigma_{-}\otimes \sigma_{+})
\label{Rmat}
\end{equation}
satisfying the Yang-Baxter relation ($\sigma_{j}$ are Pauli matrices;
$\sigma_{\pm }=(\sigma_{1}\pm i\sigma_{2})/2$).
Generators
$A,\,B,\,C,\,D$ are matrix elements of the $ L$- operator
\begin{equation}
%\label{lax}
L(u)=\left[
\matrix{{{ukA-u^{-1}k^{-1}D}\over {q-q^{-1}}}&C\cr
         B&{{ukD-u^{-1}k^{-1}A}\over {q-q^{-1}}}\cr}
\right]
\label{L}
\end{equation}
Here $u$ is the spectral parameter and $k$ is an additional parameter
(rapidity at the site). Note that the $R$- matrix is the $L$-
operator in
the spin 1/2 - representation. It is given by the same matrix
(\ref{L})
for $k=q^{1/2}$ with
elements: $\, A=q^{{1\over 2}\sigma_3},\,D=q^{-{1\over
2}\sigma_3},\,B=\sigma_{+},\,C=\sigma_{-}$.

Irreducible finite  $2j+1$ dimensional representations can be
expressed in the weight basis where $A$ and $D$ are diagonal matrices:
\,$\,A=diag\,(q^{j},...,q^{-j})$. An integer or half-integer $j$ is the spin of

the representation. The value of the Casimir
operator (\ref{casimir})
in this representation is given by the $q$- analog of $(j+1/2)^2$.
\begin{equation}
c_{0}=\left({{q^{j+1/2}-q^{-j-1/2}}\over {q-q^{-1}}}\right)^2
=[j+1/2]^2_q
\label{casimir2}
\end{equation}

The representation can be realized by difference operators acting in the space
of polynomials $\Psi(z)$ of degree $2j$ .
\begin{eqnarray}
\label{reprs}
&&A\Psi(z)=q^{-j}\Psi(qz),\,D\Psi(z)=q^{j}\Psi(q^{-1}z),
\nonumber\\
&&B\Psi(z)=z(q-q^{-1})^{-1}\left(q^{2j}\Psi(q^{-1}z)-q^{-2j}\Psi(qz)\right)
\nonumber\\
&&C\Psi(z)=-z^{-1}(q-q^{-1})^{-1}\left(\Psi(q^{-1}z)-\Psi(qz)\right)
\end{eqnarray}
Then $\Psi_0(z)=1$ is the lowest weight vector whereas $\Psi_{2j}(z)=z^{2j}$ is
the highest weight vector,i.e.
$C\Psi_0(z)=0\,,\,B\Psi_{2j}(z)=0$

We call this series of
representations "regular".They are
a smooth deformation of the representation of the $sl_2$ algebra
by differential operators:
\begin{equation}
S_3=z{d \over dz}-j,\,S_{+}=z(2j-z{d \over dz}),\, S_{-}={d \over dz}
\label{sl2}
\end{equation}

In addition, in the special dimension $Q$ there is three parametric family of
representations having, in general, no lowest and no highest weight.
Sometimes they are called cyclic or unrestricted representations.
These representations have no classical finite dimensional limit.
They can be written in the Weyl
basis \cite{bazhanov}, \cite {Sk1}, \cite{RA}. Let $X$ and $Y$ satisfy
\begin{equation}
qXY=YX
\label{Weyl}
\end{equation}
Then the representation of $U_q(sl_2)$ is
\begin{eqnarray}
A=X,\,D=X^{-1},
\nonumber\\
B=(bX+\bar b X^{-1})Y^{-1},
\nonumber\\
C=Y(cX+\bar c X^{-1}),
\label{bazhanov}
\end{eqnarray}
and is characterized by parameters $b,c,\bar b, \bar c$ obeying the conditions
$qbc=q^{-1}\bar b \bar c =-(q-q^{-1})^{-2}$.
 The value of the Casimir operator
(\ref{casimir}) depends on the parameters and is equal to
$c\bar b+\bar c b -2(q-q^{-1})^{-2}$.
 Comparing it with (\ref{casimir2}) we find
that at $q^2 b/\bar b=\bar c/c=\mp q $ (for $P$-odd, even) cyclic
representations belong to the "regular" series with
 $q^{2j+1}=\mp 1$ for $P$ - odd (even) and
the Casimir operator (\ref{casimir2})  is
\begin{eqnarray}
\label{casimir3}
&&c_{0}=-4(q-q^{-1})^{-2},\, for \,P-odd
\nonumber\\
&&c_{0}=0,\,for \,P-even
\end{eqnarray}

\section{Magnetic translations as a special representation of the
quantum group}

For a given value of the Bloch wave vector $\vec k$ the dimension of physical
Hilbert space of our problem is $Q$. A representation of dimension $2j+1=Q$
of the quantum group $U_q(sl_2)$ for $q=exp(i\pi {P\over Q})$ naturally acts in
the space of Bloch states in the magnetic field. Therefore magnetic
translations and the Hamiltonian (1),(5) can be expressed through generators of
the quantum group. Let us set
 \begin{eqnarray}
\label{XY}
&&T_y=e^{ip_y}YX^{-1},\,
\nonumber\\
&&T_x= e^{ip_x}YX,
\end{eqnarray}

Then, the Hamiltonian (\ref{ham2}) could be expressed as a linear form in
quantum group generators. At the middle of the band (an integrable point)
\begin{equation}
\label{point}
e^{i (p_x-p_y)}=- q^{-1}
\end{equation}
we may choose
\begin{eqnarray}
\label{bc}
&&b=e^{{i\over 2} (p_y-p_x)}(q-q^{-1})^{-1},
\nonumber\\
&&\bar b={q^2}( q-q^{-1})^{-1},
\nonumber\\
&&\bar c=(1-q^2)^{-1},
\nonumber\\
&&c=e^{{i\over 2} (p_y-p_x)}(1-q^2)^{-1},
\end{eqnarray}

Then $B$ and $C$ given by (\ref{bazhanov}) form a representation of the quantum
group with the  value of the Casimir operator
Eq.(\ref{casimir3}) wich corresponds to the regular representation
i.e. with the highest and the lowest weight.
Using (\ref{bazhanov}) and (\ref{XY}) we may identify the quantum group
generators and magnetic translations
\begin{eqnarray}
T_{-x}+T_{-y}=(q-q^{-1})e^{-{i\over 2}(p_y+p_x )}B,
\nonumber\\
T_{x}+T_{y}= \mp(q-q^{-1}) e^{{i\over 2} (p_y+p_x )}C,
\nonumber\\
T_{-y}T_{x}=\pm q^{-1}A^2,
\nonumber\\
T_{-x}T_{y}=\pm qD^2
\end{eqnarray}
The Hamiltonian now acquires the form
\begin{equation}
H=(q-q^{-1})(\mp e^{{i\over 2} (p_y+p_x )}C+ e^{-{i\over 2} (p_y+p_x
)}B) \label{ham31}
\end{equation}
In fact at the integrable point (\ref{point}) the physical states and their
energies ( midband states) do not depend on the value of
$\exp{{i\over 2} (p_y+p_x )}$. It simply can be gauged away.   As a result  for
the midband states in the isotropic case the relation between magnetic
translations and the quantum group is as follows \begin{equation}
 H=i(q-q^{-1})(C\pm B) \label{ham3}
\end{equation}

The Hamiltonian (\ref{ham3}) can be written as the trace of a modified $L $-
operator (\ref{L}) $\tilde
L(u)$.  Say at $P$ odd  \begin{equation}
\label{tr1}
H=Tr\tilde L(u)=Tr L(u)\sigma_{1}
\end{equation}
 Modified $L$-operator also obeys the intertwining relation (\ref{RLL=LLR})
\footnote {${}^{*}$ $L$-operator of such kind is used in
the sine-Gordon model \cite{F}. We are grateful to L.Faddeev and A.Volkov for
pointing our attention to the Ref.\cite{F}.}

Realization of the quantum
group in terms of magnetic translations is not  unique. There are different
realizations wich may be used for different physical applications. Below we
present, another "dual" realization. Say for an  odd $P$ we have:
\begin{eqnarray} \label{trqr2} &&T_{-x}+T_{-y}=-i(q-q^{-1})q^{-{1\over 2}}BD,
\nonumber\\
&&T_{x}+T_{y}=- i(q-q^{-1})q^{-{1\over 2}}CA,
\nonumber\\
&&T_{-y}T_{x}=q^{-1}A^2,\,T_{-x}T_{y}=qD^2
\end{eqnarray}
that also forms a representation of $U_q(sl_{2})$ with the same value
of the Casimir operator (\ref{casimir2}) and with the same
dimension $Q$. Now the Hamiltonian turns into quadratic form in terms ofthe
$U_q(sl_2)$ generators:
\begin{equation}
H=-i(q-q^{-1})q^{-{1\over 2}}(CA+BD)
\label{ham4}
\end{equation}
Two different forms of the Hamiltonian are in fact gauge equivalent: they
correspond to the two choices of gauge discussed in
Sect.2.
To stress ambiguity of representation of the quantum group by magnetic
translations we notice that the Hamiltonian of our problem can be also
represented as a quadratic form of another quantum group $U_{q^2}(sl_2)$
\begin{equation}
 H=CA-BD+q^2BA-q^2CD
\label{ham5}
\end{equation}
Diagonalization of this form leads to the same Bethe Ansatz equation as the
form
(\ref{ham4}).

 In the Appendix B we show that the Hamiltonian (\ref{ham4})
represented as a quadratic form of the quantum group generators can also be
understood as a trace of a quantum monodromy matrix, though of a more
complicated nature - the monodromy matrix for a system with non-periodic
boundary conditions.

 Since the main ingredients of the quantum inverse scattering method have
appeared in the problem there is a strong evidence that the Hofstadter
Hamiltonian can be actually diagonalized for arbitrary $Q$. i.e. its energy
spectrum may be expressed by solutions of the Bethe equations. The ambitious
goal of solving the problem for arbitrary parameters $p_x,p_y$ and $t_x/t_y$ is
beyond of the scope of this paper. Instead we shall present a solution for the
fixed values of the parameters when representation of the quantum group
(expressed through
magnetic translations)
belong to the "regular" series and, therefore,
 has the lowest and the highest weight. This point corresponds to the middle
of band and isotropic hoping. In this case the so-called functional Bethe
Ansatz \cite{Sk2} is applicable.

\section{functional Bethe Ansatz}

Tthe Hamiltonian of the Bloch particle in a
magnetic field  in the midband point of the spectrum $\Lambda=0$ can be
expressed as a polynomial form in the quantum group operators in a regular
representation. Let us use this advantage and replace generators by their
functional realizations.

Let us first consider the representation in which the Hamiltonian is a linear
form(\ref{ham3}).
Substituting the functional realization (\ref{reprs}) in the
 Hamiltonian (\ref{ham3}) we obtain a difference equation for a
meromorphic function $\Psi(z)$
\begin{equation}
i(z^{-1}+qz)\Psi(qz)-i(z^{-1}+q^{-1}z)\Psi(q^{-1}z)=E\Psi(z)
\label{eq1}
\end{equation}
 This equation can be easily obtained directly from the original Harper
equation in the chiral gauge (\ref{LCharper})
 without a reference to the quantum
group. Indeed, Eq.(\ref{eq1}) is an extension of the (\ref{LCharper}) at $\vec
p=({\pi \over 2}, {\pi\over2})$ to the whole complex plane provided
\begin{equation}
\label{rest1}
\chi_{n}=\Psi(q^{n})
\end{equation}
However the miracle which would be very hard to see without the quantum group
is that the energy in the extended equation (\ref{eq1}) does not depends on
$z$.

Now it is easy to get a family of Hamiltonians which have the same spectrum.
Let us set $z=q^nu$ and $\Psi(q^nu)=\chi_n(u)$. Then,
all Hamiltonians
$$H(u)=i(q^{-n}u^{-1}+q^{n+1}u)\chi_{n+1}(u)-i(q^{-n}u^{-1}
+q^{n-1}u)\chi_{n-1}(u)=E\chi_n(u)$$ have the same spectrum
and correspond to $\Lambda=0$. The original Harper's  Hamiltonian
(\ref{harper} ) is $H(1)$.

Another miracle is that we know in advance that the function $\Psi(z)$ is a
polynomial of degree $Q-1$:
 \begin{equation}
\Psi(z)=\prod_{m=1}^{Q-1} (z -z_m)
\label{polyn}
\end{equation}
Note that last fact is not valid in an arbitrary gauge, in particular
in the Landau gauge.

The similar arguments work in case of the modified Landau gauge.
 The  functional
realization of the quantum group Hamiltonian (\ref{ham4}) gives the difference
equation
\begin{equation}
z^{-1}\Psi(q^2 z)+q^{-2}z\Psi(q^{-2}z)-(z+z^{-1})\Psi(z)=E\Psi(z)
\label{eq4}
\end{equation}
with a polynomial solution. It is an extension of the Harper equation
 in the modified Landau
gauge for $\vec k'=(0,\pi )$ can be extended to the whole
complex plane with
\begin{equation}
\label{rest2}
\phi_{n}=\Psi (q^{2n})
\end{equation}

The most suitable method to solve the spectral problems (\ref{eq1},\ref{eq4})
or
 is the so called {\it functional Bethe-Ansatz}.

Let us plug (\ref{polyn}) in (\ref{eq1},\ref{eq4}) and divide  both sides
by $\Psi(z)$.\begin{eqnarray}
\label{eq13}
&&i(z^{-1}+qz)\prod_{m=1,m\ne l}^{Q-1} {{qz -z_m}\over {z- z_m}}
\nonumber\\
&&-i(z^{-1}+q^{-1}z)\prod_{m=1,m\ne l}^{Q-1} {{q^{-1}z -z_m}\over
{z-z_m}}=E
\end{eqnarray}
The l.h.s. of this equation is a meromorphic function, whereas the
r.h.s.
is a constant. To make them equal we must null all residues of the
l.h.s..
They appear at $z=0$,\,at $z=\infty$\, and at $z=z_m$.
The residue at $z=0$ vanishes automatically.

The residue at $z=\infty$\ is $-iq^Q+iq^{-Q}$. Its null determines
the
deqree of the polynom.

Comparing  the coefficients of $z^{Q-1}$ in the both sides of
Eq.(\ref{eq13}),we obtain the energy given by Eq.(\ref{energy1})
 advertised  in the Introduction.

Finally, annihilation of poles at $z=z_m$ gives the  Bethe-Ansatz
equations
(\ref{be}) for roots of the polynomial (\ref{polyn}).
  Here we write them
in a more conventional form. Let
$z_l=\exp {(2\varphi_l)}$, then
\begin{equation}
{{\cosh(2\varphi_l-i{\Phi\over 4})}\over
{\cosh(2\varphi_l+i{\Phi\over 4})}}=
\pm \prod_{m=1,m\ne l}^{Q-1}{{\sinh(\varphi_l-\varphi_m+i{\Phi\over
4})}\over
{\sinh(\varphi_l-\varphi_m-i{\Phi\over 4})}}
\label{BEconv}
\end{equation}
 The Bethe equations for the quadratic  realization of the  Hamiltonian
(\ref{ham4}),can be obtained via similar method. They are \begin{equation}
z^2_l=
q^Q\prod_{m=1,m\ne l}^{Q-1} {{q^2 z_l-z_m}\over {z_{l}-q^2
z_m}},\,\,\,\,\, l=1,...,Q-1
\label{be4}
\end{equation}
The energy spectrum is again proportional to the sum of roots:
\begin{equation}
E=-q(q-q^{-1})\sum_{l=1}^{Q-1}z_l.
\label{energy4}
\end{equation}
Inspite of the apparent difference eqs. (\ref{be}) and  (\ref{energy1})
must be equivalent to the eqs. (\ref{be4}) and (\ref{energy4}). The former is
useful tro describe the middle of the spectrum, whether the latter is good for
the bottom of the spectrum.

\section{q-Analog of Orthogonal Polynomials as Exact Zero Mode
Wave Functions}

There is an intriguing connection between the wave function of the zero energy
state  and q-generalization of the classical orthogonal polynomials  \cite{AW}.
 The q-analoq of orthogonal  polynomials (so called
Askey-Wilson polynomials) are the most general polynomials orthogonal on a
descrete support. They depends on four parameters
(except  $q$) and satisfy a q-analog of the differential hypergeometric
 equation. It is the difference equation
\begin{eqnarray}
\label{aw}
A(z)P_n(q^2z)+A(z^{-1})P_n(q^{-2}z)-(A(z)+A(z^{-1}))P_n(z)
=(q^{-2n}-1)(1-abcdq^{2n-2})P_n(z)
\end{eqnarray}
where
\begin{equation}
A(z)={{(1-az)(1-bz)(1-cz)(1-dz)}\over {(1-z^2)(1-q^2z^2)}}
\end{equation}
and $a,b,c,d$ are
parameters and $n$ is the degree of the polynomial.Note, that  $P_n$ is a
{\it Laurent} polynomial in $z$ and usual polynomial in $z+z^{-1}$ of
degree $n$. Recently, the Askey-Wilson polynomials appeared in
connection with representation of the quantum group. It has been
found that they are closely related to $6-j$ -symbol of
$U_q(sl_2)$ \cite{KR}. They also appeared in "$q$-harmonic analysis" to be
 spherical functions on the quantum $SL(2)$ group
\cite{NM}.
Some of them are closely related to our problem.

 Choosing  $c=-d=q$, $a=b=0$  we
arrive at the equation for the continuous $q$-Hermite polynomials \cite{AW}
$H_n^{(q)}$ $$
H_n^{(q)}(q^2z)-{z}^2 H_n^{(q)}(q^{-2}z)=q^{-2n}(1-{z}^2
)H_n^{(q)}(z).
$$
(They are called continuous because of their orthogonality on the
unit circle with a continuous measure).

It is clear from (\ref{eq4}) that for an odd $Q$
 $q$-Hermite polynomial of the order $n=(Q-1)/2$
yields zero energy solution of our problem:
\begin{equation}
\Psi^{(E=0)}(iz)=z^{(Q-1)/2}H_{(Q-1)/2}^{(q)}(z).
\end{equation}
The explicit form of the $q$-Hermite polynomials is
\begin{equation}
H_n^{(q)}(z)=\sum_{m=0}^{n} { {(q^{2};q^{2})_{n}}\over
{ (q^{2};q^{2})_{m}(q^{2};q^{2})_{n-m}} } z^{2m-n}
\end{equation}
where the standard notation
$$
(a;q)_{n}=\prod_{l=0}^{n-1}(1-aq^{l})
$$
is used.

Another choice is $c=-d=q$,  $a=-b=q$  and then  the replacement of
$q$ by
$q^{1/2}$
gives the q-Legendre equation
\begin{equation}
{{1-q{z}^2}\over {1-{z}^2}}P_n^{(q)}(qz)+{{q-{z}^2}\over {
1-{z}^2}}P_n^{(q)}(q^{-1}z)=(q^{-n}+q^{n+1})P_n^{(q)}(z)
\end{equation}
Comparing with the Eq.(\ref{eq1}) we conclude that the zero
mode solution is given by the continuous $q$-Legendre polynomial
  $$\Psi^{(E=0)}
(iz)=z^{(Q-1)/2}P_{(Q-1)/2}^{(q)}(z)$$
Their explicit form is
\begin{equation}
P_n^{(q)}(z)=\sum_{m=0}^{n} {
{(q;q)_{n}(q^{1/2};q)_{m}(q^{1/2};q)_{n-m}}
\over {(q;q)_{m}(q;q)_{n-m}} } z^{2m-n}
\end{equation}

\section{Conclusion}

We have showed that motion of a particle in a periodic potential in a magnetic
field has posses a structure of the quantum group. We found the Bethe Ansatz
equation for the midband spectrum and the wave function for a particle on a
square lattice. We also presented a class of quasiperiodic equations
related to $U_q(sl_2)$ which can be solved in a similar way. The most
interesting
feature of that type of equation is the multifractality of their spectrum at
$P,Q\rightarrow \infty$, i.e when flux $\Phi/2\pi$ is irrational. The spectrum
is complex but not chaotic. Quite opposite, it is determine by the quantum
integrability.

As we already mentioned the Bethe Ansatz solution is available (but not
presented here) for an arbitrary strength of the potential of the Harper
equation. It provides the basis to study Anderson localization-delocalization
transition of electrons in quasiperiodic potential.

As the best of our knowledge even elementary questions regarding the
Hofstadter problem at irrational flux remained unclear. For example  the low
temperature thermodinamics, dispersion of excitations, conductivity etc. are
not
known. In all previous examples of quantum integrable models all these
quantaties have been found by solving Bethe Ansatz equations at large $Q$.
In this limit  dispersion of bands becomes negledgible.
 Therefore, the lack of solution away from the midband is
therefore is not an obstacle for study multifractality and other physical
properties.
 The Bethe Ansatz solution does not give an advantage at finite $P$ and
$Q$. Just contrary, up to $Q\sim 6000$ direct diagonalization of the
Hofstadter Hamiltonian is more effective. However, as usual the Bethe Ansatz
solution becomes a powerful tool at $Q\rightarrow\infty$.
Perhaps the most interesting task is to solve the Bethe Ansatz equations in
this limit. The strategy of solving the Bethe equations is well known: at large
$Q$ the roots $z_l$ form dense groups ("strings") and can be described by their
distributions. The algebraic equations  then  replaced by the system of
integral
equations for the distribution functions of strings. This program is in
progress.

Another ambitious problem is to generalize  the Bethe ansatz
solution
to the whole Hilbert space of the model.

\section{Supplement A:A class of difference and discrete equations related to
the
quantum group}

A genral quadratic  form in generators of the quantum group provide
a class of difference and discrete operators of the second order which allows
complete or partial algebraization. A problem of the Bloch particle in magnetic
field is a particular case (the most physically valuable, perhaps) of the
general class.

The idea of algebraization is simple. Consider a regular
representation of the quantum group with the lowest and the highest weight. The
space of this representation can be  can be realized by the ring of polynomials
of degree $2j$. Any form in generators $A,\,B,\,C,\,D$ preserves the space of
polynomials. Therefore eigenfunctions of a form, which by means of the
representation (\ref{reprs}) is a difference operator, are polynomials as well.
Vice versa is probably also
true - one may look on the quantum group  as an algebra of difference operators
which preserve the ring of polynomials.

 Quadratic forms of  the "classical" $sl_2$ as a class of
 second-order diffferential
equations solvable in polynomials (so called "quasi
exactly solvable" models of quantum mechanics ) have been extensively studied
\cite{Turb} \cite{Ush} (see also \cite
{Shifman} for a revew) in the last few years. They include
second order differential
equations for all classical orthogonal polynomials and a set of Schrodinger
operators with solvable potentials. Attempt of q-generalization for difference
functional equations (in fact very closed to our approach) has been made in
Ref.\cite{Og} \footnote {${}^{*}$ We are indebted to A.Turbiner who informed us
about this work}.

Consider a general quadratic form (in the Appendix B we show that it can be
consider as a trace of monodromy matrix of some integrable model)
\begin{eqnarray} \label{G}
G=aA^2+dD^2+(q-q^{-1})(c_{2}CA+b_{2}BD+b_{3}BA+c_{3}CD)+
(q-q^{-1})^{2}(b_{1}B^{2}+c_{1}C^{2})
\end{eqnarray}
with a set of
parameters $a,\,d,\,c_{i},\,b_{i}$ ($i=1,2,3$).

Under the representation  (\ref{reprs}) the quadratic form is a
difference operator. Let us consider its spectral problem.
\begin{equation}
\label{diffeq}
G\Psi (z)=a(z)\Psi (q^{2}z)+d(z)\Psi (q^{-2}z)-v(z)\Psi (z)=E\Psi (z)
\end{equation}
where
\begin{equation}
\label{a(z)}
a(z)=b_{1}q^{-4j+1}z^{2}-b_{3}q^{-3j}z+aq^{-2j}+c_{2}q^{-j}z^{-1}+
c_{1}q^{-1}z^{-2}
\end{equation}
\begin{equation}
\label{d(z)}
d(z)=b_{1}q^{4j-1}z^{2}+b_{2}q^{3j}z+dq^{2j}-c_{3}q^{j}z^{-1}+
c_{1}qz^{-2}
\end{equation}
\begin{equation}
\label{v(z)}
v(z)=(q+q^{-1})(b_{1}z^{2}+c_{1}z^{-2})+(c_{2}q^{-j}-c_{3}q^{j})z^{-1}
+
(b_{2}q^{-j}-b_{3}q^{j})z
\end{equation}

If all $b_{1,2,3}=0$ the difference equation (\ref{diffeq})
 is known as $q$-hypergeometric equation (see e.g. \cite{AW}). Their solutions
are certain $q$-Jacobi  polynomials - a degenerate case of Askey Wilson
polynomials (\ref{aw}) \cite {GZ}. At the "classical" limit $q\rightarrow 1$
they
reproduce classical orthogonal polynomials. The difference equation
(\ref{diffeq}) in it general form in the classical limit leads to Mathieu and
Lame equations. One may call the equation (\ref{diffeq}) $q$- analog of the
Mathieu (Lame) equations.

{}From the theory of representation of the quantum group, we know that
this equation has $(2j+1)$ polynomial solutions of degree no greater than $2j$.

\begin{equation}
\Psi(z)=\prod_{m=1}^{N} (z-z_m)
\label{polynom00}
\end{equation}

The class of "integrable" equations can be extended if we apply
a "gauge" transformation
$$\Psi(z)\rightarrow f(z)\Psi(z),\,a(z)\rightarrow
a(z)f(z)/f(q^2z),\,d(z)\rightarrow d(z)f(z)/f(q^{-2}z),\,v(z)\rightarrow
v(z)$$.

A class of "integrable" discrete equations can be obtained from (\ref{diffeq})
by setting $z=q^{2n}, \, a(q^{2n})=a_n,\,d(q^{2n})=d_n ,\,v(q^{2n})=v_n,\,
\Psi(q^{2n})=\Psi_n$. \begin{equation}
\label{disceq}
a_n\Psi _{n+1}+d_n\Psi _{n+1}-v_n\Psi _n=E\Psi _n,\,\,n=1,....,Q
\end{equation}
Solutions of the discrete equations are given by discrete polynomials
\begin{equation}
\Psi_{n}=\prod_{m=1}^{N} (q^{2n}-z_m)
\label{polynom7}
\end{equation}
with the same roots.

The {\it functional Bethe-Ansatz} described above determines algebraic
equations for roots of the solution   (\ref{polynom00}) of eqs.
(\ref{diffeq}) or (\ref{disceq})  \footnote
{${}^{**}$. We note that this method is most close to the approach of
algebraization of quantum mechanical problem suggested by
Ushveridze\cite{Ush}}. We already applied this method for the Harper Equation.
For complete reference we repeat it for a general case below.
First of all, let us
plug (\ref{polynom00}) in (\ref{diffeq}) and divide  both sides by $\Psi(z)$.
We
get \begin{equation}
\label{res}
a(z)\prod_{m=1}^{N} {{q^{2}z -z_m}\over {z- z_m}}+
d(z)\prod_{m=1}^{N} {{q^{-2}z -z_m}\over {z-z_m}} -v(z)=E
\end{equation}
If at least one of the coefficients $c_{1},\,c_2,\,c_3$ is nonzero, there are
two different cases:

(i) at least one of $b_1,\,b_2,\,b_3$ is nonzero,

(ii) All $b's$ are zero -the quadratic form (\ref{G})
 includes only $A,\,D$ and $C$
(generators of the Borel subalgebra of $U_q(sl_2)$). That type of  difference
equation (\ref{diffeq}) is known as $q$-{\it hypergeometric equation}. Their
solutions are certain $q$-Jacobi  polynomials - a class of polynomials
orthogonal with a discrete measure (see e.g. \cite{AW}).

Let us consider the case (i) first. The l.h.s. of (\ref{res}) is a meromorphic
function, whereas the r.h.s. is a constant. To make them equal we must cancel
all the singularities of the l.h.s.
They appear at singular points of $a(z)$, $d(z)$ and $v(z)$ (double
and simple poles at $z=0$ and $z=\infty$) and at $z=z_m$.

The singular part at $z=0$ vanishes automatically.

Vanishing of the singular part at $z=\infty$\
\begin{eqnarray}
\label{singinf}
b_{1}(q^{2N-4j+1}+q^{-2N+4j-1}-q-q^{-1})z^{2}+
b_{2}(q^{-2N+3j}-q^{-j})z+b_{3}(q^{j}-q^{2N-3j})z+
\nonumber\\
+z(q-q^{-1})b_{1}(q^{2N-4j}-q^{-2N+4j})\sum_{m=1}^{N}z_m
\end{eqnarray}
determines the degree of the polynomial: $N=2j$.

Comparing  the constant terms in the both sides of (\ref{res}) we
find the energy spectrum:
\begin{equation}
\label{energy}
E=b_{1}(q-q^{-1})(q^{2}-q^{-2})\sum_{n<m}^{N}z_{n}z_{m}-
(q-q^{-1})(b_{2}q^{-j+1}+b_{3}q^{j-1})\sum_{m=1}^{N}z_m+
aq^{2j}+dq^{-2j}
\end{equation}
Finally, annihilation of poles at $z=z_m$ gives the following
Bethe-Ansatz equations
\begin{equation}
{{d(z_{l})}\over {a(z_{l})}}=
q^{4j}\prod_{m=1,m\ne l}^{2j} {{q^{2} z_l-z_m}\over {z_{l}-q^{2}
z_m}},\,\,
\,l=1,...,2j.
\label{BE1}
\end{equation}
The Bethe equations is a system of $2j$ algebraic equations. It must
 have exactly $2j+1$ solutions corresponding to different
 eigenfunctions. In the case (i)
all of them are polynomials of one and the same degree $2j$.

Similar arguments are applicable in the case (ii) . The
difference is that the l.h.s. of (\ref{res}) is now regular at $z=\infty$ from
the very beginning and the condition of vanishing of (\ref{singinf}) does not
bring a  restriction on degree of the  polynomials. The Bethe equations
 are valid
for any $N<2j+1$
\begin{equation}
{{d(z_{l})}\over {a(z_{l})}}=
q^{2N}\prod_{m=1,m\ne l}^{N} {{q^{2} z_l-z_m}\over {z_{l}-q^{2}
z_m}},\,\,
\,l=1,...,N.
\label{BE2}
\end{equation}
 Apparently, for each degree there is exactly one such polynomial. This means
that for each  $N<2j+1$ the Bethe equations (\ref{BE2}) must have exactly one
solution. This fact is far from obvious when we look at (\ref{BE2}).
 Anyway, we derived equations (\ref{BE2})  for zeros of $q$-orthogonal
polynomials. It seems to be interesting to review the theory of orthogonal
polynomials from this point of view.

The main property of the $q$-hypergeometric equation (the case (ii)) that makes
it "solvable" is triangularity of the matrix connecting the original basis
$z^{m}$\, ($m=0,1,...,2j$) with the basis formed by eigenfunctions of the
operator $G$. In particular, this leads to a very simple structure of the
spectrum of $G$:
\begin{equation}
\label{Esolv}
E_{N}=aq^{2N-2j}+dq^{2j-2N},\,\,\,\, N=0,...,2j.
\end{equation}
\\
\\

\section{supplement B:Quantum integrable models with non-periodic boundary
conditions and Hofstadter problem}

A general quadratic form in quantum group generators is related to a quantum
magnetic chain with one site and non-periodic boundary conditions.

To begin with, let us give a brief summary of the formalism treating
the integrable systems with boundaries. The boundary conditions of an
integrable model are
 determined by $c$-number $2\times 2$ matrices $K_{+}(u)$ and
$K_{-}(u)$  depending
on spectral parameter and satisfying the "reflection
equations" \cite{Cher}.
\begin{eqnarray}
\label{refl}
&&R({u/v})(K_{-}(u)\otimes 1)R(uvq^{-1})(1\otimes K_{-}(v))=
(1\otimes K_{-}(v))R(uvq^{-1})(K_{-}(u)\otimes 1)R({u/v})
\nonumber\\
&&R({v/u})(K_{+}^{t}(u)\otimes 1)R((uvq)^{-1})(1\otimes
K_{+}^{t}(v))=
(1\otimes K_{+}^{t}(v))R((uvq)^{-1})(K_{+}^{t}(v)\otimes 1)R({v/u})
\end{eqnarray}
($t$ means the transposition) with the $R$-matrix \ref{Rmat}.
 Each solution of the "reflection
equations" specify a boundary condition consistent with integrability.
Solutions for $K_{+}$ and for $K_{-}$ are related
\begin{equation}
K_{+}(u)=K_{-}^{t}(u^{-1})
\end{equation}
The monodromy matrix for an integrable model with non-periodic boundary
conditions  is a {\it  quadratic} form in the monodromy matrix $L(u)$ of a
model with periodic boundary conditions \cite{Sk3}.
\begin{equation}
\label{T}
T(u)=L(u)K_{-}(u)\sigma_{2} L^{t}(u^{-1})\sigma_{2}
\end{equation}
The trace of the monodromy matrix
\begin{equation}
\label{tau}
\tau (u)=Tr(K_{+}(u)T(u))
\end{equation}
forms a commutative family $[\tau (u),\tau (v)]=0$.

It is known that for the models with the trigonometric $R$-matrix
there is a 3-parametric family of boundary $K$-matrices \cite{Gaudin},
\cite{dV}
\begin{equation}
\label{Kminus}
K_{-}(u)=\left(
\matrix{\alpha (q^{-1}s^{-1}u-qsu^{-1})&\beta
(q^{-1}u^{2}-qu^{-2})\cr
\gamma (q^{-1}u^{2}-qu^{-2})&-\alpha (su-s^{-1}u^{-1})\cr}
\right)
\end{equation}
\begin{equation}
\label{Kplus}
K_{+}=\left(
\matrix{\lambda (qtu-q^{-1}t^{-1}u^{-1})&\mu (qu^{2}-q^{-1}u^{-2})\cr
\nu (qu^{2}-q^{-1}u^{-2})&-\lambda (t^{-1}u-tu^{-1})\cr}
\right)
\end{equation}
where $\alpha,\,\beta,\,\gamma,\,s$ and $\lambda,\,\mu,\,\nu,\,t$ are arbitrary
parameters characterized the boundary conditions. Substituting the $L$-operator
(\ref{L}) into (\ref{T}) we find matrix elements of the transfer matrix
\begin{eqnarray}
\label{T11}
&&T_{11}(u)={{\alpha (q^{-1}s^{-1}u-qsu^{-1})}\over{(q-q^{-1})^{2}}}
(k^{2}+k^{-2}-u^{2}A^{2}-u^{-2}D^{2})+
\nonumber\\
&&{{q^{-1}u^{2}-qu^{-2}}\over{q-q^{-1}}}(\gamma ku^{-1}CD-
\gamma k^{-1}uCA-\beta kuAB+\beta k^{-1}u^{-1}DB)+\alpha
(su-s^{-1}u^{-1})CB
\end{eqnarray}
\begin{eqnarray}
\label{T12}
&&T_{12}(u)={ { q^{-1} u^{2} - qu^{-2} }\over {q-q^{-1}}}\,
({{\beta (k^{2}A^{2}+k^{-2}D^{2}-u^2-u^{-2})}\over {q-q^{-1}}} -
\gamma (q-q^{-1})C^2+
\nonumber\\
&&+\alpha (sk^{-1}DC-s^{-1}kAC))
\end{eqnarray}
\begin{eqnarray}
\label{T21}
&&T_{21}(u)={{q^{-1}u^{2}-qu^{-2}}\over {q-q^{-1}}}\,
({{\gamma (k^{2}D^{2}+k^{-2}A^{2}-u^2-u^{-2})}\over {q-q^{-1}}}
-\beta (q-q^{-1})B^{2}+
\nonumber\\
&&+\alpha (skBD-s^{-1}k^{-1}BA))
\end{eqnarray}
\begin{eqnarray}
\label{T22}
&&T_{22}(u)={{\alpha (s^{-1}u^{-1}-su)}\over {(q-q^{-1})^{2}}}
(k^2+k^{-2}-u^{2}D^{2}-u^{-2}A^{2})+
\nonumber\\
&&{{q^{-1}u^{2}-qu^{-2}}\over {q-q^{-1}}}(\beta ku^{-1}BA-\beta
k^{-1}uBD-
\gamma kuDC+\gamma k^{-1}u^{-1}AC)-\alpha (q^{-1}s^{-1}u-qsu^{-1})BC
\end{eqnarray}

Summing diagonal elements (\ref{T11}) and (\ref{T22}) we find the trace of the
transfer matrix $\tau (u)$
\begin{eqnarray}
\label{tauexpl}
&&\tau (u)={{ (u^4 + u^{-4}-q^2 - q^{-2} )}\over {(q-q^{-1})^{2}}}\,
[(\mu \gamma k^{-2}+\nu \beta k^{2}-\alpha \lambda ts^{-1})A^2+
(\mu \gamma k^{2}+\nu \beta k^{-2}-\alpha \lambda t^{-1}s)D^{2}+
\nonumber\\
&&(q-q^{-1})(-(tk^{-1}\lambda \gamma +q^{-1}s^{-1}k\alpha \nu )CA+
(q^{-1}t^{-1}k^{-1}\lambda \beta +sk\alpha \mu )BD+
\nonumber\\
&&(t^{-1}k\lambda \gamma +qsk^{-1}\alpha \nu )CD-
(qtk\lambda \beta +s^{-1}k^{-1}\alpha \mu )BA)-
\nonumber\\
&&-(q-q^{-1})^{2}(\mu \beta B^{2}+\nu \gamma C^{2})\,] + c-number term
\end{eqnarray}
(here we replace  the Casimir element (\ref{casimir}) by it $c$-number value )

The trace of the transfer matrix $\tau (u)$ is a general quadratic form  in
$A,\,B,\,C,\,D$ since it depends on $7$ parameters -
$3$ in each $K$-matrix and rapidity  $k$. Indeed, the total number of
coefficients of a general quadratic form is 10 but one of them is a common
multiplier and another two contribute to the
 $c-number term$ in(\ref{tauexpl}) due to the
two central elements ($AD=1$ and the Casimir operator).

As we showed  previously  the Hamiltonian of the Bloch particle in magnetic
field is a particular quadratic form of the quantum group generators and
therefore can be considered as an integrable model.
\appendix{A}{Gauges}
1. The connection between Landau gauge and the modified Landau
 gauge (\ref{MLG}) is
very simple
\begin{equation}
\label{trans1}
\phi_{n}(\vec k')=\exp({{i\Phi }\over {2}}n(n-Q))\psi_{n}(\vec k)
\end{equation}
\begin{equation}
\label{L-ML}
 k'_{x}=k_{x}+{{i\pi P(Q-1)}\over {Q}},\,\,\,\,\,\,\,\,
 k'_{y}=k_{y}
\end{equation}
Both  $k'_{x}$ and $k_{x}$ are defined modulo $\Phi$ .

2.The relationship between Landau gauge and the chiral gauge is a bit
sofisticated \footnote {${}^{*}$}{We are indebted to A.Abanov who found this
connection}. They are connected by the Fourier transformation
\begin{equation}
\label{Fourier}
{\tilde \phi_{m}}=\sum_{n=0}^{Q-1}e^{i\Phi nm}\phi_{n}
\end{equation}
where $\phi_{n}$ is the wave function in the
 modified Landau gauge.  The new function
${\tilde \phi_{n}}$ obeys the equation
\begin{eqnarray}
\label{fourmodharper}
&& (e^{ik'_{y}}+e^{i\Phi n- i k'_{x}}){\tilde \phi_{n+1}} +
\nonumber\\
&&+(e^{-i k'_{y}}+ e^{-i\Phi n +i\Phi
+i k'_{x}}){\tilde \phi_{n-1}}
=E{\tilde \phi_{n}}
\end{eqnarray}
Comparing with (\ref{modLCharper}) we can identify the momenta as follows:
\begin{equation}
 k'_{x}=-p_{y}-{1\over 4}\Phi,\,\,\,\,\,\,\,\,
 k'_{y}=p_{x}-{1\over 4}\Phi
\end{equation}
\noindent
{\bf ACKNOWLEDGMENTS}\noindent
\\
We would like to thank  P.G.O.Freund, A.Abanov, A.Gorsky, A.Kirillov,
E.Floratos, J.-L.Gervais and V.Pasquier for interesting
discussions.

P.W. acknowledges the hospitality of Weizmann Institute of Science
and Laboratoire de Physique Theorique de l'Ecole Normale
Sup\'{e}rieure where part of this work was done. A.Z. is grateful to
the Mathematical Disciplines
Center of the University of Chicago for the hospitality and support.
He also would like to thank Prof. A.Niemi for the warm hospitality
in the Institute of Theoretical Physics at Uppsala University where
this work was completed. This work was supported in part by  NSF
under the Research  Grant 27STC-9120000 and NSF-DMR 88-19860
\newpage
Postscriptum: when this paper has been finished we have learned that L.Faddev
and R.Kashaev succeded in extending the Bethe Ansatz solution for an arbitrary
momentum and an arbitrary anisotropy. We indebt to L.Faddeev and R.Kashaev for
informing us of their results.

%\end{narrowtext}

\newpage



\begin{references}

\bibitem{Z} J.Zak, {\sl Phys. Rev. } {\bf134} 1602
(1964)

\bibitem {A} M.Ya.Azbel, {\sl Sov. Phys.JETP} {\bf19} 634 (1964)

\bibitem{W} G.H.Wannier,{\sl Phys.Status Solidi} {\bf 88}, 757 (1978)
\bibitem{H} D.R.Hofstadter,{\sl Phys. Rev. B} {\bf 14 } 2239 (1976)

\bibitem{TKNN} D.J.Thouless, M.Kohmoto, P.Nightingale, M.den Nijs,
{\sl
Phys.Rev.Lett} {\bf49},405 (1982)

\bibitem{AA} S.Aubry and G.Andre {\sl  Ann.Israel Phys.Soc.} {\bf 3},
131
(1980)

\bibitem{T} D.J.Thouless, {\sl Phys. Rev.} {\bf28} 4272 (1983)

\bibitem{K} H.Hiramoto, M.Kohmoto, {\sl Int.J.Mod.Phys. B} {\bf 6},
281
(1992)
\bibitem{Tak} M.Takahashi and M.Suzuki , {\sl Prog. Theor.Phys. } {\bf48} 2187
(1972)
\bibitem{Jap} G.I.Japaridze, A.A.Nersesyan and P.B.Wiegmann {\sl Nucl.Phys.  }
{\bf B230} 511 (1984)

\bibitem{Anderson} P.W.Anderson in  Proc. of Nobel Symp. 73 {\sl
Physica
Scripta} {\bf T27}, (1988)

\bibitem{wiegmann} P.B.Wiegmann  in  Proc. of Nobel Symp. 73 {\sl
Physica
Scripta} {\bf T27}, (1988)
\bibitem{wieg} P.B.Wiegmann  invited talk in Nobel Symp. 73, Sweden (1988),
unpublished.
\bibitem{KR} P.P.Kulish,  N.Yu.Reshetikhin {\sl Zap.nauch.semin.LOMI}
{\bf
101}, 112 (1980)
\bibitem{Zab} P.B.Wiegman and A.Zabrodin. Algebraizaition of a class ofdiscrete
equations related to $U_q(sl_2)$.
\bibitem{Dr} V.G.Drinfeld , {\sl Dokl.Acad.Nauk } {\bf 283} 1060
(1985)
\bibitem{J} M.Jimbo {\sl Lett.Math. Phys.} {\bf 10} 63 (1985)

\bibitem{FRT} N.Yu.Reshetikhin, L.A.Takhtadjan, L.D.Faddeev,{\sl
Algebra i Analiz} {\bf 1}, 178 (1989)

\bibitem{RA} P.Roche, D.Arnaudon {\sl Lett.Math. Phys.} {\bf 17}, 295
(1989).

\bibitem{Sk1} E.K.Sklyanin {\sl Func.Anal.Appl.} {\bf 17}, 273 (1983)

\bibitem{Fl} E.G.Floratos {\sl Phys.Lett .} {\bf B233}, 235 (1989)

\bibitem{F} L.D.Faddeev {\it The Bethe Ansatz}, Andrejewski lectures,
preprint SFB 288, No 70 (1993)

\bibitem{Sk2}  E.K.Sklyanin  {\sl Zap.nauch.semin.LOMI} {\bf 134},
112
(1983)
\bibitem{KR} A.Kirilov and  N.Yu.Reshetikhin {\sl Infinite dimensional Lie
Algebra and Groups,Advanced Study in Math.Phys. v.7,ed.V.Kac,p.285,World
Scientific 1988}
\bibitem{Turb} A.V.Turbiner {\sl Comm.Math. Phys.} {\bf 118} 467
(1988)

\bibitem{Ush}  A.Ushveridze {\sl Sov.Journal Part.Nucl.} {\bf 20},
185
(1989), {\it ibid} {\bf 23}, 25 (1992)

\bibitem{Cher}  I.Cherednik {\sl Theor.Mat. Fys.} {\bf 61} 35 (1984)

\bibitem{Sk3}  E.K.Sklyanin {\sl  J.Phys. A} {\bf 21}. 2375 (1988)

\bibitem{bazhanov}  V.V.Bazhanov and Yu.G.Stroganov {\sl  J.Stat.Phys.} {\bf
59}. 799 (1990)

\bibitem{Gaudin}  M.Gaudin {\it La Fonction d'Onde de Bethe}, Masson, 1983

\bibitem{dV}  H.J. de Vega, A.Gonzalez-Ruiz {\sl J.Phys. A} {\bf 26}
L519  (1993)

\bibitem{Shifman}  M.A.Shifman {\sl  Int.Journal of Mod.Phys.A} {\bf 4} 2897
(1989)

 \bibitem{Og} O.V.Ogievetsky and A.V.Turbiner{\sl   Preprint
CERN-TH}:6212/91 (1991)

 \bibitem{AW}  see e.g.G.Gasper and M.Rahman, {\it Basic
Hypergeometric
Seriaes} (1990), Cambridge Univ.Press  and N.Vilenkin, A.Klimyk, {\it
Representation of Lie Groups and Special Functions},vol.3 (1992)
Kluwer
Acad.Publ
\bibitem{NM}M.Noumi and K.Mimachi  {\sl Proc. Japan Acad. Ser. A} {\bf 66} 146
(1990)

\bibitem{GZ} A general Askey-Wilson polynomial can be obtained in a similar way
from a degenerate Sklyanin algebra :
A.Gorsky and A.Zabrodin {\sl J.Phys. A} {\bf 26} L635  (1993)

\end{references}
\end{document}